# Crossover from magnetostatic to exchange coupling in La$_{0.67}$Ca$_{0.33}$MnO$_3$/YBa$_2$Cu$_3$O$_7$/La$_{0.67}$Ca$_{0.33}$MnO$_3$ heterostructures

Rajni Porwal[1,2], Anurag Gupta[1,2] and R C Budhani[3]

[1] Academy of Scientific and Innovative Research, CSIR-NPL, New Delhi-110012, India
[2] CSIR-National Physical Laboratory, New Delhi-110012, India
[3] Condensed Matter-Low Dimensional Systems Laboratory, Indian Institute of Technology Kanpur, Kanpur-208016, India

E-mail: rcb@iitk.ac.in

**Abstract**

The influence of YBa$_2$Cu$_3$O$_7$ (YBCO) superconductor layer (S-layer) with varying thickness $d_{YBCO}$ = 20 to 50 nm on the magnetic coupling between two La$_{0.67}$Ca$_{0.33}$MnO$_3$ (LCMO) ferromagnet layers (F-layer, thickness $d_{LCMO}$ = 50 nm) in F/S/F heterostructures (HS's) was investigated by measuring global magnetization ($M$) in a temperature ($T$) range = 2 – 300 K and magnetic field ($H$) range = 0 – 10 kOe. All the HS's were superconducting with critical temperature ($T_C$) decreasing from = 78 to 36 K with decrease in $d_{YBCO}$, whereas the ferromagnetic ordering temperature $T_M$ = 250 K did not change much. Systematically measured M-H loops of all HS's at both $T > T_C$ and $T < T_C$ show three main results: (a) the two step magnetic reversal above $T_C$ converts into a four step reversal below $T_C$ in HS's with $d_{YBCO} \geq$ 30 nm, (b) the magnetic field corresponding to the additional two switching steps and their magnitude show characteristic evolution with $T$ and $d_{YBCO}$; and (c) the HS with $d_{YBCO}$ = 20 nm shows radically different behaviour, where the two step magnetic reversal above $T_C$ continues to persist below $T_C$ and converts into a single step reversal at $T \ll T_C$. The first two results indicate magnetostatic coupling between the magnetic domains and the vortices across the two F/S interfaces resulting in reversal dynamics different from that deep within the LCMO layers. Whereas, the result 'c' reveals indirect exchange coupling between LCMO layers through the superconducting YBCO layer, which is a clear experimental evidence of coexistence of ferromagnetism and superconductivity in nm scale F/S/F HS's expected theoretically by C.A.R. Sa de Melo (Physica C 387, 17-25 (2003)).

# 1. Introduction

F/S/F heterostructures (where F and S refer to ferromagnet and superconductor, respectively) have and continue to attract much attention due to the interesting physics arising from various couplings like proximity, magnetostatic, and exchange at the S/F interface [1-4]. One of the recently pursued system has been based on cuprate high $T_C$ (i.e., superconducting critical temperature) superconductor (CHTS) and mangnanite colossal magnetoresistance (CMR) materials, which have been considered special due to several reasons: very similar perovskite related crystalline structures of both with nearly matching lattice parameters that help grows sharp S/F interface; CMR being half metals and fully spin polarized have smaller exchange fields; and CHTS have small coherence lengths. A variety of novel phenomena like spin triplet superconductivity [5-8], superconducting spin switch [9-12], and vortex pinning by magnetic domain wall interactions [13] have been reported in them. Besides, the $T_C$ modulation of the S-layer in the presence of F-layer, the contrasting magnetic nature of S/F interface and deep inside the F-layer has also interested the researchers [14, 15]. The later has been investigated by techniques such as neutron reflectometry and XMCD [14, 15]. Another interesting case of coexistence of magnetism and superconductivity was raised early on theoretically by C. A. R. Sa de Melo [16] in nanometer scale CMR/CHTS/CMR HS. He suggested a situation where without destroying the superconductivity and magnetism completely in the HS, the S-layer leads to exchange coupling between the F-layers.

We address these issues through detailed global magnetization reversal measurements of the La$_{0.67}$Ca$_{0.33}$MnO$_3$ (LCMO) F-layers of fixed thickness ($d_{LCMO}$ = 50 nm) and YBa$_2$Cu$_3$O$_7$ (YBCO) S-layer with varying thickness ($d_{YBCO}$ = 50 to 20 nm) in F/S/F geometry. Since a magnetic exchange coupling between the F-layers is expected to exist for a S-layer in normal state only when its thickness is less than ~ 13 nm [17], in all our samples both the LCMO layers are expected to be magnetically decoupled at $T > T_C$. As reported in refs. [18, 19], choice of $d_{YBCO} \geq 20$ nm ensures that YBCO layer will retain superconductivity in all our samples. The carefully chosen geometry and dimensions thus allow explicit investigations of the effect of superconducting YBCO layer on the decoupled LCMO layers. At $T < T_C$, indeed a strong influence of YBCO layer, where, depending on its thickness both magnetostatic coupling at the S/F interface and indirect exchange coupling between F-layers are found to play a dominant role in the magnetic reversal dynamics of the LCMO layers. The significant physics that emerges from our work is the magnetic coupling between the two F-layers in the presence of intervening superconducting condensate in our HS samples.

## 2. Experimental details

The epitaxial thin films of $La_{0.67}Ca_{0.33}MnO_3/YBa_2Cu_3O_7/La_{0.67}Ca_{0.33}MnO_3/LaCrO_3$ were deposited on (001) $LaAlO_3$ (LAO) substrates using multitarget pulsed laser deposition based on KrF excimer laser ($\lambda$ = 248 nm) [20]. Since LCMO (a = b = c = 3.868 Å in bulk) has an excellent lattice matching with $LaCrO_3$ (LCO) (a = b = c = 3.885 Å in bulk) [21], the latter was grown on LAO substrate to provide a buffer layer for an epitaxial growth of the former. The thickness of the spacer YBCO layer ($d_{YBCO}$) was varied from 20 to 50 nm, whereas the same for the LCMO ($d_{LCMO}$) and LCO ($d_{LCO}$) layers was kept constant at 50 nm in all the hetrostructures (HS's). Depending on the $d_{YBCO}$ = 50, 40, 30, and 20 nm, these samples henceforth are named as HS-50, HS-40, HS-30 and HS-20, respectively. The structural studies were performed by using X-ray diffraction measurement with Cu-K$_\alpha$ radiation ($\lambda$ = 1.5405 Å). All the samples used for both transport and magnetization measurements had dimensions of 2 x 5 mm$^2$. Four probe transport measurements in temperature range $T$ = 5 - 310 K were done in a customized close-cycle refrigerator setup. The magnetization measurements were carried out in a SQUID-based magnetometer (MPMS-XL7, Quantum Design) with field applied in in-plane orientation for all the samples in temperature range $T = 2 - 300$ K and magnetic field range B = 0 - 10 kOe.

## 3. Results and Discussion

Figure 1 shows the $\theta$-$2\theta$ X-ray diffraction profiles of all the four samples (HS-50, HS-40, HS-30 and HS-20) where the $d_{YBCO}$ changes from 50 to 20 nm. The observed (00$l$) reflections of LCO, LCMO and YBCO confirm the highly oriented growth of all the compounds. With the increasing $d_{YBCO}$ the (005)-YBCO peak becomes narrower and intense, as also reported in ref. [22].

The superconducting and magnetic transitions in all the HS samples were established by both transport and magnetization measurements. Figure 2(a) displays the resistivity $\rho$(T) of all the four samples (HS- 50, 40, 30 and 20). The $T_C$ is defined as the temperature at which the resistance of the sample drops to zero in the $\rho$(T) curve, which is marked by arrow in the inset of Figure 2(a). The $T_M$ is defined by the crossing point of two lines deduced from the linear region of the insulating state and the metallic state in the $\rho$(T) curve, as indicated by arrow in the Figure 2(a). For magnetically measured superconducting transition, the samples were initially cooled in a field of 10 kOe (to ensure fully saturated state of the LCMO layer) until $T \sim 100$ K, below which the zero field cooled (ZFC) measurements of magnetization $M$(T) were performed in a 100 Oe field applied in in-plane orientation, as shown in the Figure 2(b). The characteristic onset of superconductivity ($T_C$) in the YBCO layer of all the samples is marked by small kinks in $M$(T) which is shown in Figure 2(a) and upper inset of Figure 2(b). For measuring the magnetic transition of the LCMO layers, the field cooled (FC) magnetization was carried out in in-plane field of 100 Oe. A clear upturn of $M$(T) marks the characteristic paramagnetic to ferromagnetic transition ($T_M$) in all the samples, shown as superimposed graphs in lower inset of Figure 2(b). We plot both $T_M$ & $T_C$ as a function of $d_{YBCO}$ in Figures 3(a) and 3(b), respectively. As seen in the Figure 3(a), the values of $T_M$ measured both resistively (RT) and magnetically (FC/ZFC) are in good agreement. $T_M$ values are nearly independent of $d_{YBCO}$ >30 nm and show a tendency to increase below 30 nm. In case of $T_C$, the values obtained magnetically are slightly lower as compare to those obtained resistively (Figure 3(b)). This is expected due to the fact that the former measurements are a bulk effect. $T_C$ decreases monotonically with decrease in $d_{YBCO}$, whereas the decrease becomes rapid at lower values of $d_{YBCO}$. It is interesting to note that $T_M$ is nearly independent of $d_{YBCO}$, whereas $T_C$ decreases rapidly at $d_{YBCO} \leq 30$ nm. A similar behaviour of $T_C$ suppression in LCMO/YBCO bilayers was reported in refs. [18, 19] and explained by spin polarized electrons diffusion, with an estimated length ~ 10 nm, in YBCO-layer.

The magnetic hysteresis loops of all the HS samples (HS-50, HS-40, HS-30 and HS-20) were measured both below and above the $T_C$. To fully saturate the LCMO layers the samples were field cooled in $H = 10$ kOe in in-plane orientation down to various $T \leq 120$ K at which isothermal M-H loops were recorded in a field range of ±5 kOe. Figure 4(a) shows the M-H loops for HS-50 sample ($T_C = 78$ K) at various $T$. For $T$ ($>T_C$) = 120 and 80 K, when the YBCO layer is in the normal state, at high positive fields the magnetization of both LCMO layers are in a saturated and parallel state. On reversing the field magnetic reversal occurs in two steps presumably due to the two decoupled LCMO layers. In between the two steps, the "plateau" in the M-H loop indicates an antiparallel state of both the layers. Such M-H plateaus are well known and responsible for the superconducting spin valve effect in F/S/F trilayers [10]. The switching fields at both the steps are estimated from the derivative of left arm of the M-H loop, for instance at 120 K see the upper inset of Figure 4(a). The peaks in the derivative at $H_{11}$ (~180 Oe) and $H_{21}$ (~295 Oe) mark the switching fields of the two steps. The fact that our bilayer LCMO/LCO film ($d_{LCMO} = 50$ nm) showed a magnetic reversal ~300 Oe (data not shown) indicates that $H_{21}$ should correspond to the bottom LCMO layer of our HS. We thus conclude that the top LCMO layer reverses first, followed by the bottom LCMO layer. At 40 K ($<T_C$), when the YBCO layer is in the superconducting state, the two step magnetic reversal of the HS converts in to four step reversal, where additional two new sharp switching steps appear (see Figure 4(a)). The four switching fields are estimated by the observed peaks in the derivative of the M-H loop, as shown for the left arm at 40 K in the lower inset of Figure 4(a). These characteristic fields are marked as $H_{11}$, $H_{12}$, $H_{21}$ and $H_{22}$, where $H_{12}$ and $H_{22}$ mark the sharp steps that appeared below $T_C$ in the M-H loop. The use of the same nomenclature $H_{11}$ and $H_{21}$ as above for relatively two broad steps out of the four will be justified later. Note also that, in contrast to 80 and 120 K, the M-H loop at 40 K shows irreversible behaviour at high positive and negative fields due to the superconducting YBCO layer.

The observation of four stepped magnetic reversal in our HS can be understood by taking into account the role of the LCMO/YBCO interface. The sandwiched YBCO layer makes two interfaces with top and bottom LCMO layers, as depicted schematically in Figure 4(b). The observed four switching steps (marked as A - D in Figure 4(a)) can be attributed to the complex magnetic dynamics in the four magnetic entities (see Figure 4(b)) in the HS, which are (i) A - top LCMO layer, (ii) B - 1st interface of the top LCMO and spacer YBCO layers, (iii) C - 2nd interface of bottom LCMO and spacer YBCO layers, and (iv) D - bottom LCMO layer. The overall M-H behaviour observed in Figure 4(a) at 40 K can now be understood as follows. In the positive saturated state, the magnetization of all the four magnetic entities is in the same direction as of the field. With decrease in magnetic field towards zero and subsequent increase in the negative direction, the first relatively broad magnetic switching step 'A' (at field $H_{11}$) reflects a partial reversal due to the magnetic domains deep inside the top LCMO layer (region marked as 'A' in Figure 4(b)). The reversal of the magnetic domains in the vicinity of the 1st F/S interface (region marked as 'B' in Figure 4(b)) gets delayed because of the vortex pinning in the adjoining YBCO layer. With an increase of field in the negative direction the sharp step 'B' (at field $H_{12}$) marks the reversal in the "B" region, when the magnetostatic energy related to magnetic domain motion wins over the vortex pinning energy. Similar delayed magnetic reversal was reported previously in a bilayer F/S system [23]. With further increase of the field in the −ve direction, the third switching step 'C' (at field $H_{22}$) is observed. The sharpness of the step indicates that it corresponds to the vortex pinning dependent reversal of the magnetized domains in the 2$^{nd}$ F/S interface (region marked as 'C' in Figure 4 (b)). Note that both the S/F interfaces show reversals at different magnetic fields, which is probable due to difference in vortex pinning and orientation of magnetic domains at the two interfaces as one moves away from the LAO substrate [24]. The complete reversal of the bottom LCMO layer (region marked as 'D' in Figure 4 (b)) occurs at higher negative fields as marked by the broader switching step 'D' (at field $H_{21}$). Finally, further increase of the field in the negative direction leads the entire HS in the negative saturated state. The measurements of the right arm results in a symmetric M-H loop with magnitudes of $H_{11}$, $H_{12}$ and $H_{21}$ that match within ±2.5 Oe, and $H_{22}$ within ±15 Oe of the left arm. We would like to mention here that the intervening "antiparallel state" is still visible at $T$ ($<T_C$) = 40 K after the top LCMO layer is reversed completely (see "plateau region" between steps B and C in Figure 4(a)).

In view of the above discussion, we now quantify the evolution of magnitude and corresponding switching field of the magnetic switching steps as a function of temperature and YBCO layer thickness in various HS's. Figures 5(a) - 5(d) show the left arm of M-H loops measured at different $T$ between 2 - 120 K for HS-50, HS-40, HS-30 and HS-20 samples, respectively. The HS's with $d_{YBCO} \geq 30$ nm (see Figures 5(a) - 5(c)) show the double step switching at all $T > T_C$, which converts into four steps at all $T < T_C$, as already discussed above for HS-50 sample. Interestingly, both the height and characteristic fields of the four steps show typical variation with temperature depending on $d_{YBCO}$. Since HS-20 sample ($d_{YBCO}$ = 20 nm) shows a totally different M-H behaviour at $T < T_C$, it will be discussed later. Figures 6(a) - 6(c) show the temperature dependence of four switching step magnitudes (see steps A, B, C, and D in Figure 4a) for HS-50, HS-40 and HS-30 samples, respectively. It might be useful to compare the magnitude of these switching steps with the expected magnetization change due to the magnetic reversal of one LCMO layer (when YBCO layer is in normal state). Assuming that both top and bottom LCMO layers having same thickness will show nearly the same magnitude, it can be taken as half of the magnetization change observed in the measured M-H loop of the HS just above $T_C$. For every HS sample, this is shown as dotted lines in the Figures 6(a) - 6(c). We observe that just below $T_C$, the magnitudes of the broad switching steps 'A' and 'D' corresponding to the top and bottom LCMO layers, respectively, nearly match with the expected one above $T_C$. Interestingly, decrease in temperature results in a decrease of the magnitude of both these steps in all three cases. This result reflects shrinkage of the regions A and D (see Figure 4(b)) responsible for partial reversal of deep regions (away from the S/F interfaces) in the top and bottom LCMO layers in HS, respectively. On the other hand, the magnitudes of the sharp switching steps 'B' and 'C' corresponding to the top and bottom S/F interfaces are nearly zero just below $T_C$. Decrease in temperature increases the magnitude of both these sharp steps, in all three cases, indicating the growth of the pinning controlled magnetic regions B and C (see Figure 4(b)) adjacent to S/F interface in the LCMO layers. The increase of vortex pinning with decrease in $T$ is evident from Figure 6(d), where relative $M_R/M_{R120}$ (where $M_R$ is the remanant state moment and $M_{R120}$ is the same at 120 K) is plotted as a function of $T$ for the four HS samples. All the samples show a pronounced increase in $M_R/M_{R120}$ below $T_C$, revealing the onset of irreversible magnetization due to vortex pinning in YBCO layer in addition to the saturated moment of the LCMO layers. At all $T(< T_C)$, with decrease in YBCO layer thickness, the values of $M_R/M_{R120}$ decrease rapidly indicative of weakening of superconductivity and vortex pinning in thinner S-layers. Surprisingly, in HS-50 sample at $T \leq 20$ K, the magnitudes of the sharp switching steps 'B' and 'C' marking the partial reversal of S/F interface region becomes much larger than that expected form the magnetic reversal of the complete LCMO layer (compare points of B and C curves with dashed line at $T \leq 20$ K in Figure 6(a)). Note that these steps cannot occur purely due to the irreversibility of the S-layer, which will always show a smoothly changing $M(H)$. This anomalous behaviour is unexpected and cannot be reconciled without invoking generation of stray field component perpendicular to the S-layer at the S/F interface. This may happen due to magnetostatic coupling between the strongly pinned vortices and magnetic domains at the S/F interface of our HS-50 sample. The results reported in refs. [3, 24] do support our conjecture.

We now look at the evolution of characteristic fields $H_{11}$, $H_{12}$, $H_{21}$ and $H_{22}$, corresponding to various switching steps for all the HS's (see Figures 5(a) - 5(d), as a function of temperature for different $d_{YBCO}$ in Figures 7(a) - 7(d). First we discuss the results of the samples HS-50, HS-40, and HS-30. For all these samples the switching fields denoted as $H_{11}$ and $H_{21}$ increasing linearly as a function of $T$ are found to lie on same continuous line both above and below $T_C$. This indicates that the switching steps pertaining to $H_{11}$ and $H_{21}$ corresponds to the magnetic regions having similar magnetic dynamics both above and below the $T_C$ in both the LCMO layers. Obviously, these regions refer to the entire LCMO layers at $T > T_C$ and only the regions deep inside the LCMO layers away from the S/F interfaces at $T < T_C$. The complete reversal of the LCMO layers mediated by two sharp steps at switching fields $H_{12}$ and $H_{22}$ originate just below $T_C$ and reveal a clear correlation with superconductivity of YBCO layer. As seen further in Figures 7(a) - 7(c), $H_{12}(T)$ shows a linear $T$ dependence for all $d_{YBCO}$, whereas $H_{22}(T)$ behaviour depends on $d_{YBCO}$. For instance, the HS-50 ($d_{YBCO}$ = 50 nm) sample initially shows a linear increase followed by a rapid increase at lower temperatures of $H_{22}(T)$. The result $H_{22} > H_{21}$ at $T \leq 20$ K for this sample (see Figure 7(a)) is a reflection of large increase of vortex pinning at the bottom S/F interface, which as discussed earlier was also responsible for the anomalously large magnitudes of the related sharp switching steps (see Figure 6(a)). In case of HS-40 sample such a change is relatively suppressed (see Figure 7(b)), whereas in HS-30 sample $H_{22}(T)$ shows a behaviour similar to $H_{12}(T)$ (see Figure 7(c)). These results agree with the earlier conclusion of decrease in vortex pinning with decrease in $d_{YBCO}$.

Now we will discuss the HS-20 sample with lowest $d_{YBCO} = 20$ nm, which shows a radically different M-H behaviour in comparison to all other samples with $d_{YBCO} \geq 30$ nm. As seen from Figure 5(d), in this sample two step switching observed at $T > T_C$ instead of converting into four step switching continues as two step at $T < T_C$ and further crosses over to single step switching at $T << T_C$. As can be appreciated from Figure 7(d), both the switching fields $H_{11}(T)$ and $H_{21}(T)$ above $T_C$ continue to persist below $T_C$ and start converging as we decrease the $T$. Further decrease of $T \leq 5K$ results in a merger of $H_{11}(T)$ and $H_{21}(T)$ on a single curve marking the single step switching field. The latter result reveals that at $T \leq 5K$ both the LCMO layers switch at same fields and the entire HS behaves like a single magnetic entity. This quantitative change in the behaviour of the HS-20 reveals that in this regime of $d_{YBCO}$, the superconducting YBCO layer indirectly exchange couples both the LCMO layers, which was absent when YBCO layer was normal. This is interestingly the same coexistence of superconductivity and magnetism in nanometer scale F/S/F structure Car Se de Malo pointed out in Ref. [16]. One possibility for magnetic coupling across a superconductor (which is YBCO in our case) can be related to the existence of virtual quasiparticles in a d-wave superconductor [16]. This result in HS-20 ($d_{YBCO} = 20$ nm) sample clearly points out the indirect exchange interactions winning over the vortex pinning controlled magnetic switching in the S/F interface regions.

## 4. Summary

In summary, we performed detailed study of the magnetization reversal in LCMO and YBCO based F/S/F heterostructures, with varying $d_{YBCO}$ = 20 to 50 nm and constant $d_{LCMO}$, both above and below the $T_C$. For all HS's with $d_{YBCO} \geq 30$ nm, at $T < T_C$, the magnetostatic coupling between the magnetic domains in the LCMO layer and the superconducting vortices in the YBCO layer at the S/F interfaces led to two additional magnetization reversal steps. Whereas, the HS with $d_{YBCO}$ = 20 nm showed clear evidence of coexistence of ferromagnetism and superconductivity as demonstrated by conversion of two step into one step reversal of the exchange coupled HS at $T \leq 5K$ (<< $T_C$ = 36 K).

**Acknowledgments**

The authors (from CSIR-NPL) thank the Director, CSIR–NPL, for his encouragement. Rajni Porwal acknowledges the financial support from the Indo-French project (grant no. GAP 123132). The Department of Science and Technology, India is acknowledged for funding (grant no. SR/NM/NS-129/2010(G)). R.C.B. acknowledges the J C Bose National Fellowship of the Department of Science and Technology, India.

**Figure captions:**

Figure 1. X-ray $\theta$ - $2\theta$ diffraction pattern of HS-20, HS-30, HS-40, and HS-50 samples. Position of the reflections due to YBCO and LCMO are marked with arrows.

Figure 2. (a) Resistivity plot of HS-20, HS-30, HS-40, and HS-50 samples. Inset shows an enlarged view near $T_C$ in logarithmic scale. (b) Main figure and its inset show the results of temperature dependent ZFC and FC magnetization, respectively in the same samples. $T_C$ and $T_M$ values were marked by arrows in the respective figures.

Figure 3. Variation of (a) $T_M$ and (b) $T_C$ as a function of $d_{YBCO}$ of all the HS's samples (HS-20, HS-30, HS-40, and HS-50) measured resistively (RT) and magnetically (FC/ZFC).

Figure 4. (a) M-H loops at temperature 40, 80, and 120 K for HS-50 ($d_{YBCO}$ = 50 nm) sample. Insets show the left arm of hysteresis loop with its derivative at temperature of 120 K (upper panel) and 40 K (lower panel) where the peak positions are marked by arrows. (b) Schematic diagram of all the four magnetic entities taking part in over all switching dynamics of the HS.

Figure 5. A comparative view of the M-H behavior of the left arms of (a) HS-50, (b) HS-40 nm, (c) HS-30, and (d) HS-20 samples, respectively in the temperature range from 2 – 120 K.

Figure 6. Temperature dependence of the magnitudes A, B, C, and D of four switching steps in the (a) HS-50, (b) HS-40 nm, and (c) HS-30 samples, respectively (d) $M_R / MR_{(120\ K)}$ versus temperatures for the same samples where the corresponding $T_C$ are marked by arrows.

Figure 7. Magnetic switching fields $H_{11}$, $H_{12}$, $H_{21}$ and $H_{22}$ as a function of temperature extracted from the left arm of M-H loop of the (a) HS-50, (b) HS-40 nm, (c) HS-30, and (d) HS-20 samples, respectively, where the corresponding $T_C$ are marked by dotted lines.

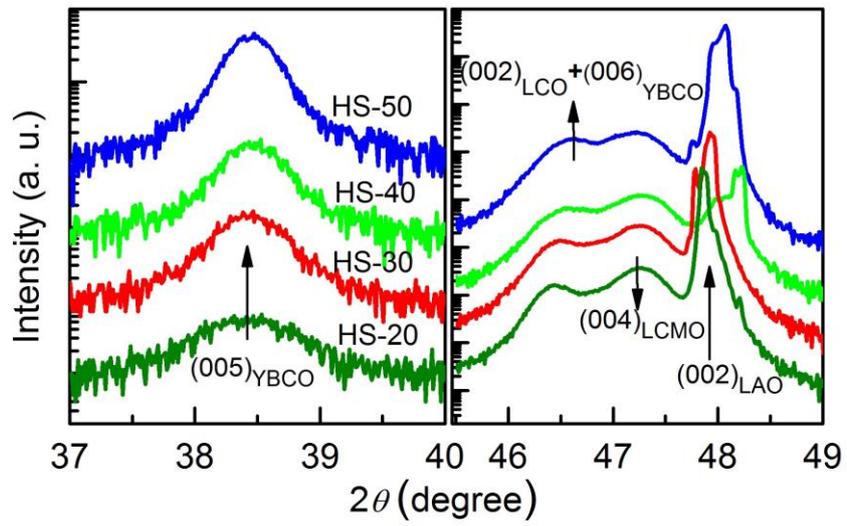

Figure 1.

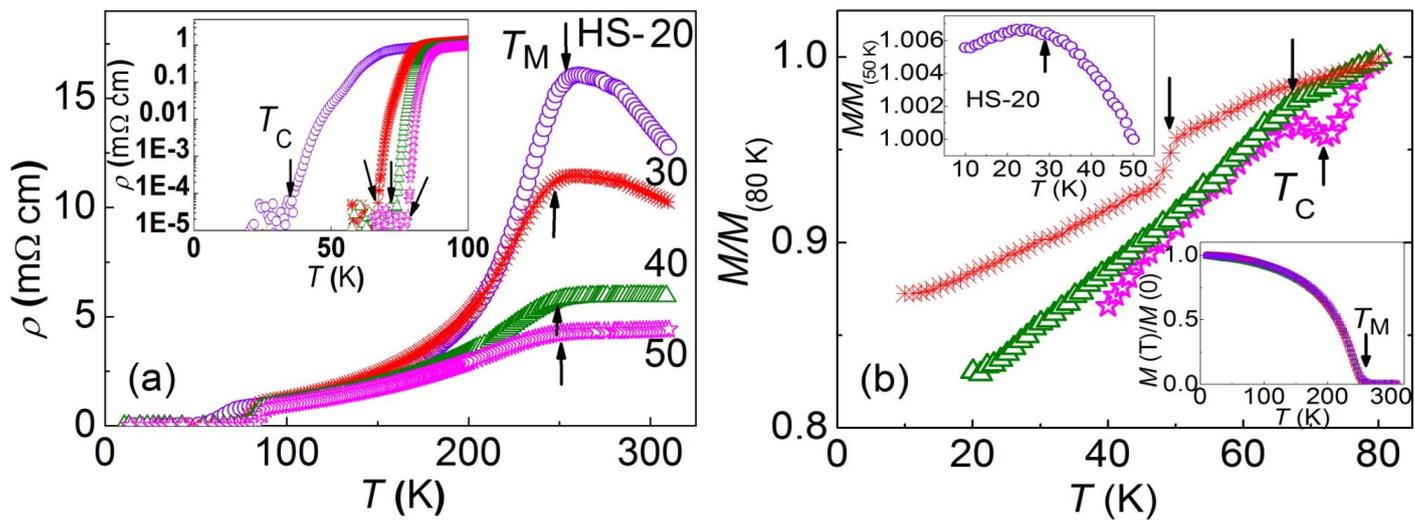

Figure 2.

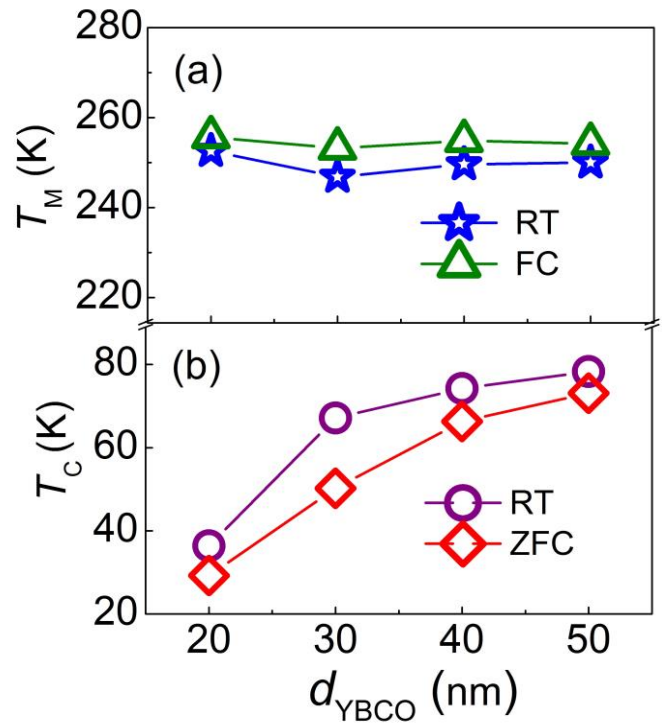

Figure 3.

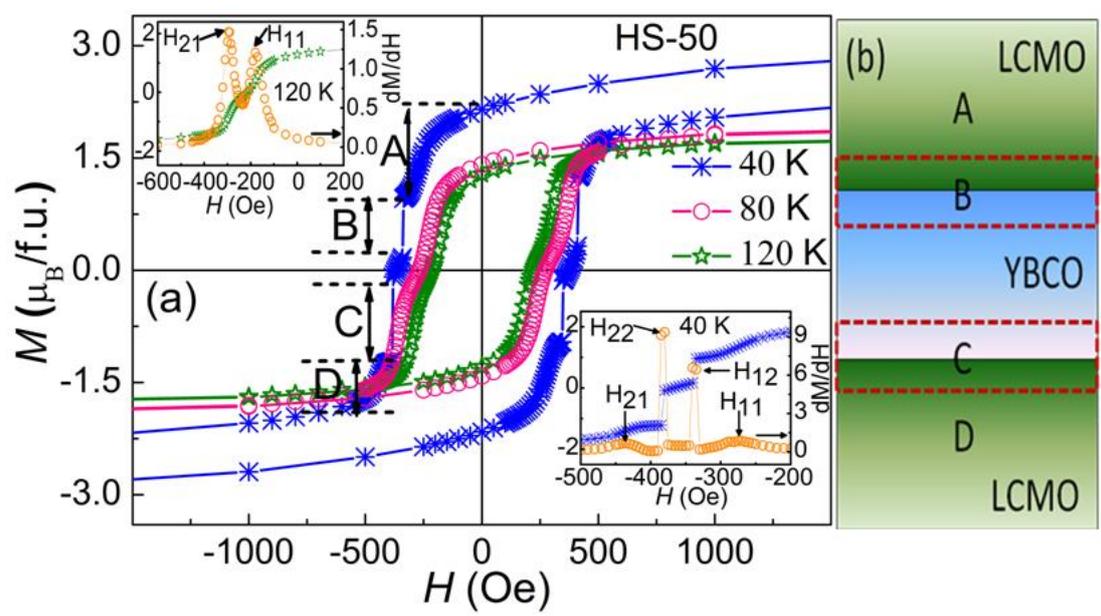

Figure 4.

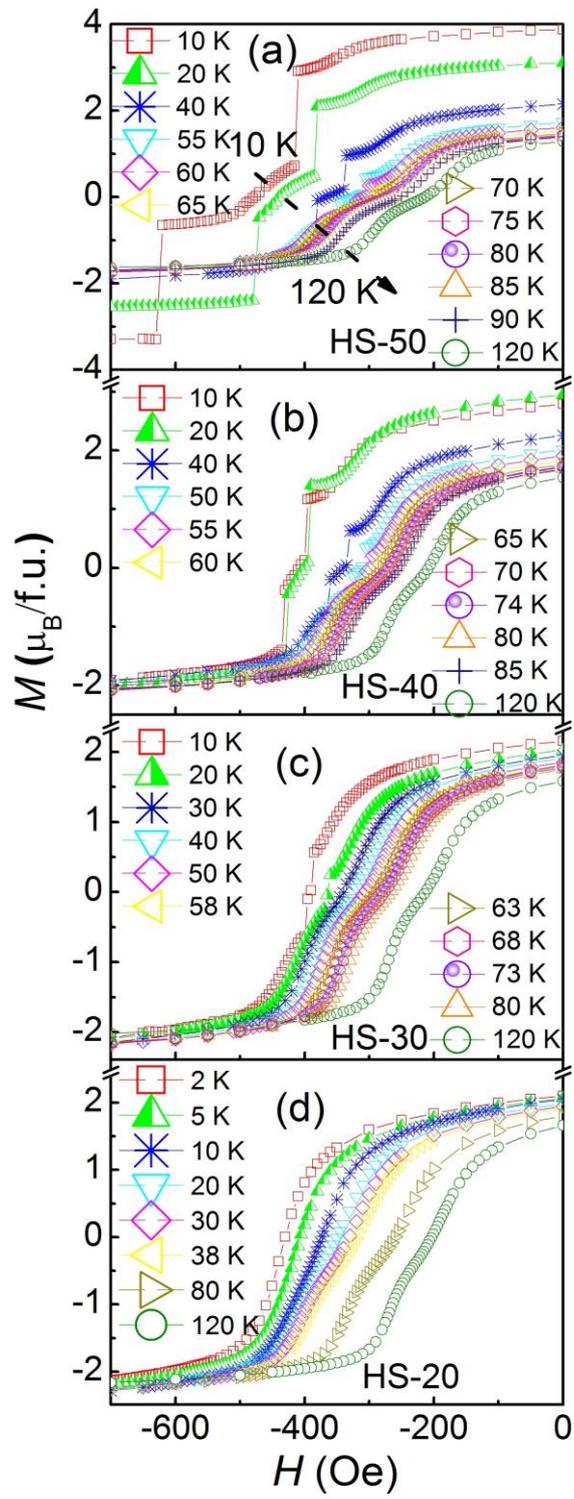

Figure 5.

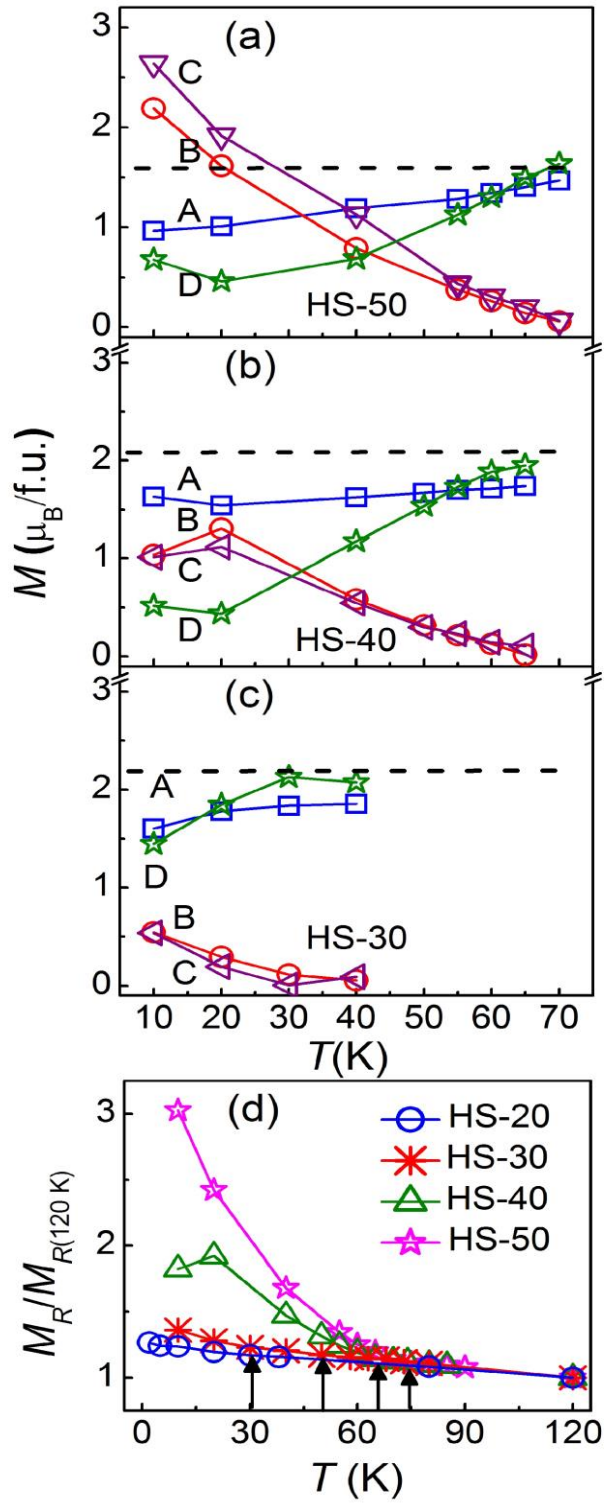

Figure 6.

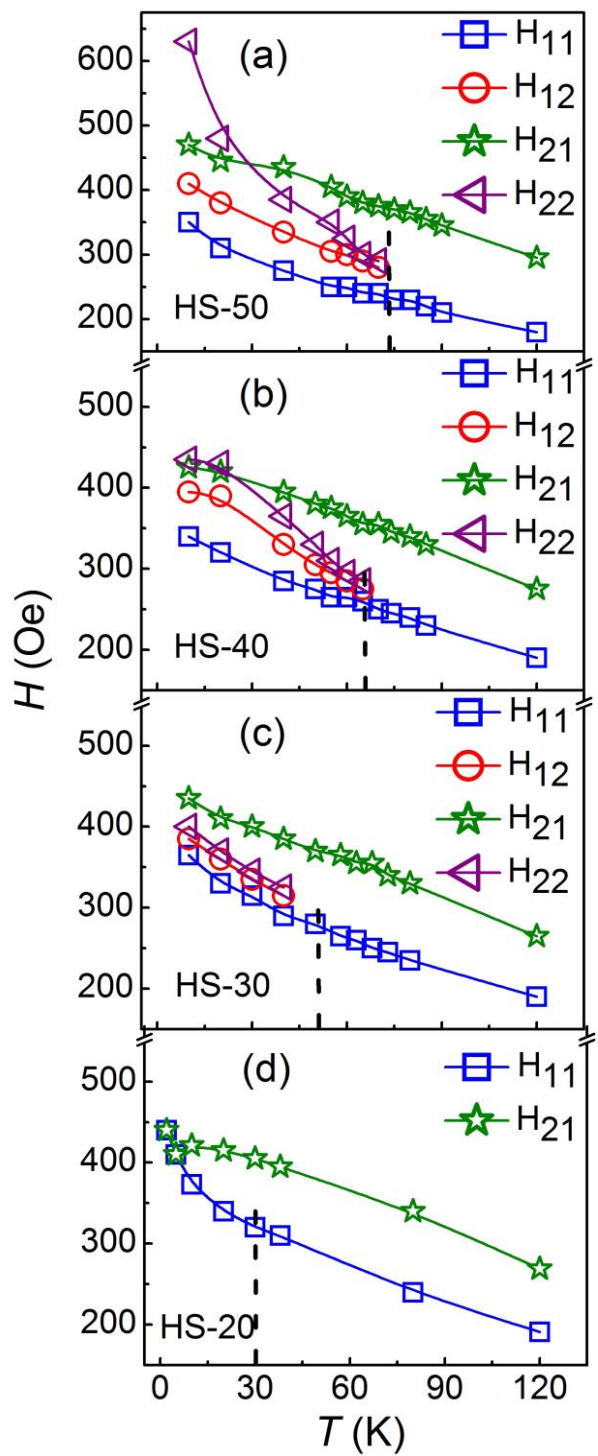

Figure 7.